\documentclass[a4paper,10pt]{article}
\usepackage{stmaryrd}
\usepackage{amsfonts}
\usepackage{bbm}
\usepackage{amscd}
\usepackage{mathrsfs}
\usepackage{latexsym,amssymb,amsmath,amscd,amscd,amsthm,amsxtra,xypic}
\usepackage[dvips]{graphicx}
\usepackage[utf8]{inputenc}
\usepackage[T1]{fontenc}
\usepackage{lmodern}
\usepackage{amssymb}
\usepackage[all]{xy}
\usepackage{nicefrac,mathtools,enumitem}
\usepackage{microtype}

\newtheorem{thm}{Theorem}[section]
\newtheorem{lem}[thm]{Lemma}
\newtheorem{cor}[thm]{Corollary}
\newtheorem{pro}[thm]{Proposition}
\newtheorem{ex}[thm]{Example}
\newtheorem{rmk}[thm]{Remark}
\newtheorem{defi}[thm]{Definition}

\newcommand{\be }{\begin{equation}}
\newcommand{\ee }{\end{equation}}

\newcommand{\pf}{\noindent{\bf Proof.}\ }

\newcommand{\HH}{\mathbb H}

\newcommand{\huaF}{\mathcal{F}}

\newcommand{\huaV}{\mathcal{V}}
\newcommand{\huaW}{\mathcal{W}}

\newcommand{\frkg}{\mathfrak g}

\newcommand{\frkG}{\mathfrak G}

 \newcommand{\R}{\mathbb R}

\def\qed{\hfill ~\vrule height6pt width6pt depth0pt}

\newcommand{\half}{\frac{1}{2}}

\newcommand{\pair}[1]{\left\langle #1\right\rangle}

\newcommand{\Courant}[1]{\left\llbracket  #1\right\rrbracket }
\newcommand{\Dorfman}[1]{\{ #1\}}

\newcommand{\jet}{\mathfrak{J}}

\newcommand{\dev}{\mathfrak{D}}

\newcommand{\Id}{\rm{Id}}

\newcommand{\dM}{\mathrm{d}}

\newcommand{\Hom}{\mathrm{Hom}}
\newcommand{\Der}{\mathrm{Der}}

\newcommand{\gl}{\mathfrak {gl}}

\newcommand{\so}{\mathfrak {so}}
\newcommand{\Symm}{\mathrm {Symm}}

\newcommand{\Ker}{\mathrm{Ker}}

\newcommand{\End}{\mathrm{End}}
\newcommand{\ad}{\mathrm{ad}}
\newcommand{\pr}{\mathrm{pr}}

\newcommand{\Img}{\mathrm{Im}}
\newcommand{\ve}{\mathrm{v}}
\newcommand{\Vect}{\mathrm{Vect}}
\newcommand{\sgn}{\mathrm{sgn}}
\newcommand{\Ksgn}{\mathrm{Ksgn}}

\newcommand{\V}{\mathbb{V}}
\newcommand{\W}{\mathbb{W}}
\newcommand{\K}{\mathbb{K}}
\newcommand{\LL}{\mathbb{L}}
\newcommand{\D}{\mathbb{D}}

\newcommand{\dperp}{{\D^0}}

\begin{document}
\title{
{Omni-Lie $2$-algebras and their Dirac structures
\thanks
 {
Research partially supported by China Postdoctoral Science
Foundation (20090451267), NSF of China (10871007), US-China CMR
Noncommutative Geometry (10911120391/A0109) and the German Research
Foundation (Deutsche Forschungsgemeinschaft (DFG)) through the
Institutional Strategy of the University of G\"ottingen.
 }
} }
\author{Yunhe Sheng  \\
Mathematics School $\&$ Institute of Jilin University,\\
 Changchun 130012,  China
\\\vspace{3mm}
email: shengyh@jlu.edu.cn\\
Zhangju Liu\\
Department of Mathematics, Peking University, \\Beijing 100871,
China\\\vspace{3mm}
email:liuzj@pku.edu.cn\\
Chenchang Zhu\\
Courant Research Center ``higher order structures'',\\
Georg-August-University
G$\rm\ddot{o}$ttingen, \\Bunsenstrasse 3-5, 37073, G$\rm\ddot{o}$ttingen, Germany\\
email:zhu@uni-math.gwdg.de }

\date{\today}


\maketitle
\begin{abstract}
We introduce the notion of omni-Lie $2$-algebra, which is a
categorification of  Weinstein's omni-Lie algebras. We prove that
there is a one-to-one correspondence between strict Lie $2$-algebra
structures on  $2$-sub-vector spaces of a $2$-vector space $\V$ and
Dirac structures on the omni-Lie $2$-algebra $ \gl(\V)\oplus \V $.
In particular, strict Lie $2$-algebra structures on $\V$ itself
one-to-one correspond to Dirac structures of the form of graphs.
Finally,  we introduce the notion of twisted omni-Lie $2$-algebra to
describe (non-strict) Lie $2$-algebra structures. Dirac structures
of a twisted omni-Lie $2$-algebra correspond to certain (non-strict)
Lie $2$-algebra structures, which include string Lie $2$-algebra
structures.
\end{abstract}

\setcounter{tocdepth}{2}

\tableofcontents

\section{Introduction}
The notion of omni-Lie algebra was introduced by Weinstein in
\cite{weinstein:omni}  to characterize Lie algebra structures on a
vector space $V$. An omni-Lie algebra can be regarded as the
linearization of the  Courant algebroid
\cite{lwx,Roytenbergphdthesis} structure on $TM\oplus T^*M$ at a
point, where $M$ is a finite dimensional differential manifold.  It
is studied from several aspects recently
\cite{clomni,kinyon-weinstein,shengzhu2,shengzhu1,UchinoOmni}. An
omni-Lie algebra associated to a vector space $V$ is the direct sum
space $\gl(V)\oplus V$ together with the nondegenerate symmetric
pairing $\pair{\cdot,\cdot}$ and the skew-symmetric bracket
operation $\Courant{\cdot,\cdot}$  given by
\begin{eqnarray*}
\pair{A+u,B+v}&=&\half(Av+Bu),\end{eqnarray*} and
\begin{eqnarray*}
\Courant{A+u,B+v}&=&[A,B]+\half(Av-Bu).
\end{eqnarray*}
With the factor of $\half$, the bracket $\Courant{\cdot,\cdot}$ does
not satisfy the Jacobi identity. However, this bracket can be
completed to a structure of a Lie $2$-algebra as in
\cite{shengzhu1}. Moreover, this Lie $2$-algebra is integrated to a
Lie 2-group  in \cite{shengzhu2}. Thus this integration procedure
provides another solution to the integration problem of omni-Lie
algebras, studied by Kinyon and Weinstein in
\cite{kinyon-weinstein}.

Notice that even though the motivating example of Courant bracket
involves an infinite dimensional vector space $\chi(M) \oplus
\Omega^1(M)$, Weinstein's linearization makes it possible to study a
finite dimensional model, namely $\gl(V)\oplus V$, where $V$, as the
tangent space $T_mM$ at certain point $m\in M$,  is a finite
dimensional vector space. Thus in our paper, we also restrict
ourselves to the finite dimensional case. That is,  all the vector
spaces in this paper are finite dimensional.

In \cite{clomni}, the authors introduced the notion of omni-Lie
algebroid to characterize Lie algebroid structures on a vector
bundle $E$. Omni-Lie algebroids are generalizations of Weinstein's
omni-Lie algebras from vector spaces to vector bundles. An omni-Lie
algebroid is the direct sum bundle $\dev E\oplus \jet E$ together
with an $E$-valued pairing and a bracket operation, where $\dev E$
and $ \jet E$ are the covariant differential operator bundle and the
first jet bundle of $E$ respectively. The main result is that Lie
algebroid structures on $E$ one-to-one correspond to Dirac
structures of the form of graphs. Moreover, (general) Dirac
structures one-to-one corresponds to projective Lie algebroid
structures on sub-vector bundles of $E\oplus TM$ \cite{clsdirac}.

Recently, people have payed more attention to higher categorical
structures with motivations from string theory. One way to provide
higher categorical structures is by categorifying existing
mathematical concepts. One of the simplest higher structure is a
$2$-vector space, which is a categorified  vector space. If we
further put Lie algebra structures on $2$-vector spaces, then we
obtain the notion of Lie $2$-algebras \cite{baez:2algebras}. The
Jacobi identity is replaced by a natural transformation, called
Jacobiator, which also satisfies some coherence laws of its own. One
of the motivating examples is the differentiation of Witten's string
Lie 2-group $String(n)$, which is called a string Lie $2$-algebra.
As $SO(n)$ is the connected part of  $O(n)$ and $Spin(n)$ is the
simply connected cover of $SO(n)$,
 $String(n)$ is a ``cover'' of $Spin(n)$ which has trivial $\pi_3$ (notice that $\pi_2(G)=0$ for any Lie group $G$). The differentiation of $String(n)$ is not
 any more  $\so (n)$, but a central extension of $\so (n)$ by the abelian Lie $2$-algebra  $\R \to 0$, which is a Lie $2$-algebra by
  itself\footnote{The concept of string Lie $2$-algebra is later generalized to any such extension of a semisimple Lie algebra.}.

To provide a way to characterize Lie $2$-algebra structures on a
$2$-vector space, we categorify Weinstein's omni-Lie algebra
$\gl(V)\oplus V$ associated to a vector space $V$. The result is the
so-called omni-Lie $2$-algebra (Definition \ref{defi:omni2})
$\gl(\V) \oplus \V$ associated to a $2$-vector space $\V$. We prove
that there is a one-to-one correspondence between Dirac structures
of the omni-Lie $2$-algebra $\gl(\V)\oplus\V$ and strict Lie
$2$-algebra structures on $2$-sub-vector spaces of $\V$. We also
introduce the notion of $\mu$-twisted omni-Lie $2$-algebra $\gl(\V)
\oplus_\mu \V$ twisted   by an isomorphism $\mu$ from $\gl(\V)$ to
itself. Dirac structures of the twisted omni-Lie $2$-algebra
$\gl(\V)\oplus_\mu\V$ characterize those Lie $2$-algebra structures
on $\V$ whose Jacobiators are determined in a specific way by the
brackets. We further verify that an interesting class of Lie
$2$-algebras including string Lie $2$-algebras is characterized by
Dirac structures.

The paper is organized as following: In Section 2, we recall some
necessary background knowledge. We construct the strict Lie
$2$-algebra $\gl(\V)$ for a $2$-vector space $\V$, which plays the
role of $\gl(V)$ in the classical case for a vector space $V$. In
Section 3, we introduce the notion of omni-Lie $2$-algebra
associated to a $2$-vector space $\V$. An omni-Lie $2$-algebra is
the $2$-vector space $\gl(\V)\oplus \V$ together with some algebraic
structures. We prove that Dirac structures of the omni-Lie
$2$-algebra $\gl(\V)\oplus \V$ characterize strict Lie $2$-algebra
structures on $2$-sub-vector spaces of $\V$ (Theorem
\ref{Thm:graphdirac}, \ref{thm:general Dirac}).  As an application
of our theory, in Section 4,  we introduce the notion of normalizer
of a $2$-sub-vector space of $\gl(\V)\oplus \V$. We prove that the
normalizer of a Dirac structure $\LL$ is a sub-Lie $2$-algebra of
$\gl(\V)$ and it can be considered as the derivation Lie $2$-algebra
(a la Schlessinger-Stasheff and Stevenson) of the Lie $2$-algebra
corresponding to $\LL$.
 In Section 5, we introduce the notion of $\mu$-twisted omni-Lie
$2$-algebras by an automorphism $\mu$ of $ \gl(\V)$. We give the
relation between Dirac structures of a $\mu$-twisted omni-Lie
$2$-algebra $\gl(\V)\oplus_\mu\V$ and  Lie $2$-algebra structures on
the $2$-vector space $\V$ (Theorem \ref{thm:twist}). Finally, we
give a description of Dirac structures corresponding to String Lie
$2$-algebras with a suitable choice of automorphism $\mu$.

\vspace{2mm}

{\bf Acknowledgement:} We give warmest thanks to Ping Xu for very
useful comments.

\section{ Lie $2$-algebras}
Vector spaces can be categorified to $2$-vector spaces. A good
introduction for this subject is \cite{baez:2algebras}. Let $\Vect$
be the category of vector spaces.

\begin{defi}{\rm\cite{baez:2algebras}}
A $2$-vector space is a category in the category $\Vect$.
\end{defi}

Thus a $2$-vector space $C$ is a category with a vector space of
objects $C_0$ and a vector space of morphisms $C_1$, such that all
the structure maps are linear. Let $s,t:C_1\longrightarrow C_0$ be
the source and target maps respectively. Let $\cdot_\ve$ be the
composition of morphisms.

 A {\em $2$-sub-vector space} of
$C$ is a $2$-vector space $C^\prime$ of which the set of morphisms
$C_1^\prime$ is a sub-vector space of $C_1$, the set of objects
$C_0^\prime$ is a sub-vector space of $C_0$, and all the structure
maps are the restrictions of the corresponding structure maps of
$C$.

It is well known that the 2-category of $2$-vector spaces is
equivalent to the 2-category of 2-term complexes of vector spaces.
Roughly speaking, given a $2$-vector space $C$,
$\Ker(s)\stackrel{t}{\longrightarrow}C_0$ is a 2-term complex.
Conversely, any 2-term complex of vector spaces
$V_1\stackrel{\dM}{\longrightarrow}V_0$ gives rise to a $2$-vector
space of which the set of objects is $V_0$, the set of morphisms is
$V_0\oplus V_1$, the source map $s$ is given by $s(v+m)=v$, and the
target map $t$ is given by $t(v+m)=v+\dM m$, where $v\in V_0,~m\in
V_1.$ We denote the $2$-vector space associated to the 2-term
complex of vector spaces $V_1\stackrel{\dM}{\longrightarrow}V_0$ by
$\mathbb{V}$:
\begin{equation}\label{v}
\mathbb{V}=\begin{array}{c}
 \mathbb{V}_1:=V_0\oplus V_1\\
\vcenter{\rlap{s }}~\Big\downarrow\Big\downarrow\vcenter{\rlap{t }}\\
 \mathbb{V}_0:=V_0.
 \end{array}\end{equation}

\begin{defi}\label{defi:Lie2}
A  Lie $2$-algebra is a $2$-vector space $C$ equipped with
\begin{itemize}
\item[$\bullet$] a skew-symmetric bilinear functor, the bracket, $[\cdot,\cdot]:C\times C\longrightarrow
C$,
\item[$\bullet$] a skew-symmetric trilinear natural isomorphism, the
Jacobiator,
$$
J_{x,y,z}:[[x,y],z]\longrightarrow [x,[y,z]]+[[x,z],y],
$$
\end{itemize}
such that the following Jacobiator identity is satisfied,
\begin{eqnarray*}
&J_{[w,x],y,z}([J_{w,x,z},y]+1)(J_{w,[x,z],y}+J_{[w,z],x,y}+J_{w,x,[y,z]})\\&=[J_{w,x,y},z](J_{[w,y],x,z}+J_{w,[x,y],z})([J_{w,y,z},x]+1)([w,J_{x,y,z}]+1).
\end{eqnarray*}
\end{defi}

A Lie $2$-algebra is called {\bf strict} if the Jacobiator is the
identity isomorphism.

\begin{rmk}This notion is called a semistrict Lie $2$-algebra in \cite{baez:2algebras}. A Lie $2$-algebra in their paper should be skew-symmetric also
up to a natural transformation. However, since we will not use this
other notion, we make this simplification, which also coincides with
Henriques' definition \cite{henriques} of Lie $2$-algebras via
$L_\infty$-algebras. The relation between Courant algebroids and
$L_\infty$-algebras is studied in \cite{rw}.
\end{rmk}

\begin{defi}\label{defi:Linfty}
An $L_\infty$-algebra is a graded  vector space $L=L_0\oplus
L_1\oplus\cdots$ equipped with a system $\{l_k|~1\leq k<\infty\}$ of
linear maps $l_k:\wedge^kL\longrightarrow L$ with degree\footnote{ the exterior powers are interpreted in the
graded sense.}
$\deg(l_k)=k-2$, such that the following relation with Koszul sign ``Ksgn'' is
satisfied for all $n\geq0$:
\begin{equation}
\sum_{i+j=n+1}(-1)^{i(j-1)}\sum_{\sigma}\sgn(\sigma)\Ksgn(\sigma)l_j(l_i(x_{\sigma(1)},\cdots,x_{\sigma(i)}),x_{\sigma(i+1)},\cdots,x_{\sigma(n)})=0.
\end{equation}
 The summation in this equation is taken over all $(i,n-i)$-unshuffles with
$1\leq i\leq n-1$.
\end{defi}
In particular, if the $k$-ary brackets are zero for all $k>2$, we
recover the usual notion of {\bf differential graded Lie algebras}
(DGLA). If $L$ is concentrated in degrees $<n$, $L$ is called an
 {\bf $n$-term $L_\infty$-algebra}.

It is well known that the notion of  Lie $2$-algebra is  equivalent
to that of  2-term $L_\infty$-algebra. In particular, strict Lie
$2$-algebras are the same as 2-term differential graded Lie algebras
(DGLA), or equivalently, crossed modules of Lie algebras. Given a
strict Lie $2$-algebra $C$, the corresponding 2-term DGLA is given
by $\ker(s)\stackrel{t}{\longrightarrow} C_0$. Conversely, given a
2-term DGLA $V_1\stackrel{\dM}{\longrightarrow}V_0$, the underlying
$2$-vector space of the corresponding strict Lie $2$-algebra is
given by (\ref{v}). Moreover, the skew-symmetric bracket  is given
by
\begin{equation}
[u+m,v+n]=[u,v]+[u,n]+[m,v]+[m,n]_\dM, \quad\forall~u,v\in
V_0,~m,n\in V_1,
\end{equation}
where the bracket $[\cdot,\cdot]_\dM$ on $V_1$ is defined by
\begin{equation}
[m,n]_\dM\triangleq[\dM m,n].
\end{equation}

Given a $2$-vector space $\V$, we 
define $\End^0_\dM(\V)$ by
$$
\End^0_\dM(\V)\triangleq\{(A_0,A_1)\in\gl(V_0)\oplus
\gl(V_1)|A_0\circ\dM=\dM\circ A_1\},
$$
and define $\End^1(\V)\triangleq \End(V_0,V_1)$. Then we have,

\begin{lem} $\End^0_\dM(\V)$ is the space of linear functors from
  $\mathbb{V}$ to   $\mathbb{V}$.
 \end{lem}
\pf Let  $(f_1,f_0)$ be a linear functor. Then $f_1$, written in the
form of a matrix of linear morphisms $V_{0,1}\to V_{0,1}$, has the
following form,
$$
f_1=\Big(\begin{array}{cc} A_0&B\\0&A_1\end{array}\Big).
$$
Therefore, for $u\in V_1$ and $m\in V_0$,  we have
\begin{eqnarray*}
s\circ f_1(u,m)=s(A_0u+Bm,A_1m)=A_0u+Bm.
\end{eqnarray*}
On the other hand, we have
$$
f_0\circ s(u,m)=f_0(u).
$$
Since a linear functor commutes with the source map and $(u,m)$ is
arbitrary, we have
$$
f_0=A_0,\quad B=0.
$$
Furthermore, we have
\begin{eqnarray*}
tf_1(u,m)=t(A_0u,A_1m)=A_0u+\dM\circ A_1m,
\end{eqnarray*}
and
$$
f_0\circ t(u,m)=f_0(u+\dM m)=A_0u+A_0\circ\dM m.
$$
By the condition that a linear functor commutes with the target map, we
have
$$
A_0\circ\dM=\dM\circ A_1.
$$
So any linear functor is of the form $ (\Big(\begin{array}{cc}
A_0&0\\0&A_1\end{array}\Big),A_0) $, where $A_0\circ\dM=\dM\circ
A_1.$

Furthermore, it is not hard to see that any linear map $
(\Big(\begin{array}{cc} A_0&0\\0&A_1\end{array}\Big),A_0) $, where
$A_0\circ\dM=\dM\circ A_1,$ preserves the identity morphisms and the
composition of morphisms. Thus $\End^0_\dM(\V)$ is the space of
linear functors from $\mathbb{V}$ to $\mathbb{V}$. \qed\vspace{3mm}

There is a differential $\delta:\End^1(\V)\longrightarrow
\End^0_\dM(\V)$ given by
$$
\delta(\phi)\triangleq\phi\circ\dM+\dM\circ\phi,\quad\forall~\phi\in\End^1(\V),
$$
and a bracket operation $[\cdot,\cdot]$ given by the graded
commutator. More precisely,  for any $A=(A_0,A_1),B=(B_0,B_1)\in \End^0_\dM(\V)$
and $\phi\in\End^1(\V)$, $[\cdot,\cdot]$ is given by
\begin{eqnarray*}
  [A,B]=A\circ B-B\circ A=(A_0\circ B_0-B_0\circ A_0,A_1\circ B_1-B_1\circ
  A_1),
\end{eqnarray*}
and
\begin{equation}\label{representation}
  ~[A,\phi]=A\circ \phi-\phi\circ A=A_1\circ \phi-\phi\circ A_0.
\end{equation}
These two operations make $\End^1(\V)\xrightarrow{\delta}
\End^0_\dM(\V)$ into a 2-term DGLA (proved in \cite{shengzhu2}),
which we denote by $\End(\V)$. This DGLA plays an important role in
the theory of representations of higher Lie algebras in
\cite{lada-markl}.  The corresponding (strict) Lie $2$-algebra of
this 2-term DGLA, denoted by $\gl(\V)$, is given by
\begin{equation}\label{glv}
\gl(\V)=\begin{array}{c}
 \End^0_\dM(\V)\oplus\End^1(\V)\\
\vcenter{\rlap{s }}~\Big\downarrow\Big\downarrow\vcenter{\rlap{t }}\\
 \End^0_\dM(\V).
 \end{array}\end{equation}
For any $A\in\End^0_\dM(\V),~\phi\in\End^1(\V)$, the source and the
target maps are given by
$$
s(A+\phi)=A\quad \mbox{and}\quad t(A+\phi)=A+\delta\phi;
$$
the skew-symmetric  bilinear functor
$[\cdot,\cdot]:\gl(\V)\times\gl(\V)\longrightarrow\gl(\V)$ is given
by
\begin{eqnarray*}
~[A+\phi,B+\psi]&=&[A,B]+[\phi,\psi]_\delta+[A,\psi]+[\phi, B],
\end{eqnarray*}
where $[\phi,\psi]_\delta$ is given by
$$
[\phi,\psi]_\delta=[\delta\phi,\psi]=\phi\circ\dM\circ\psi-\psi\circ\dM\circ\phi.
$$

The Lie $2$-algebra $\gl(\V)$ plays the same role of $\gl(V)$ in the
classical case of a vector space $V$. Another example with a similar
flavor is X. Zhu's Lie $2$-algebra $\gl(\mathcal {C})$ for any
abelian category $\mathcal {C}$ \cite{zhuxw}.  \vspace{3mm}

The Lie $2$-algebra $\gl(\V)$ acts on $\V$ naturally:
\begin{equation}\label{eqn:action}
(A+\phi)(u+m)=A(u+m)+\phi(u+\dM
m),\quad\forall~A\in\End^0_\dM(\V),\phi\in\End^1(\V).
\end{equation}
It is not hard to see that this action is a bilinear functor from
$\gl(\V)\times\V$ to $\V$. 

 The action (\ref{eqn:action}) is a generalization of the usual
 representation of $\gl(V)$ on a vector
space $V$. There is a natural Lie algebra structure on $\gl(V)\oplus
V$ which is the semidirect product of $\gl(V)$ and $V$. Similarly,
for a $2$-vector space $\V$, there is also a similar semidirect
product  strict Lie $2$-algebra structure on $\gl(\V)\oplus \V$.
This fact is proved in the case of 2-term $L_\infty$-algebras in
\cite{shengzhu2}. However, in the next section, we introduce another
bracket on  $\gl(\V)\oplus \V$ which does not make it into a Lie
$2$-algebra.

\section{Dirac structures of omni-Lie $2$-algebras}

On the direct sum $\gl(\V)\oplus\V$, we can define a $\V$-valued
nondegenerate symmetric pairing $\pair{\cdot,\cdot}$. On the space of
morphisms, it is given by
\begin{equation}\label{defi:pair1}
\pair{A+\phi+u+m,B+\psi+v+n}\triangleq\half\big((A+\phi\big)
(v+n)+\big(B+\psi) (u+m)\big).
\end{equation}
On the space of objects, it is given by
\begin{equation}\label{defi:pair0}
\pair{A+u,B+v}\triangleq\half(Av+Bu).
\end{equation}

\begin{lem}\label{lem:pair}
The $\V$-valued nondegenerate symmetric pairing $\pair{\cdot,\cdot}$
defined by \eqref{defi:pair1} and \eqref{defi:pair0} is a bilinear
functor from $(\gl(\V)\oplus\V)\times(\gl(\V)\oplus\V)$ to $\V$.
\end{lem}
\pf Obviously, the pairing
$\pair{\cdot,\cdot}:(\gl(\V)\oplus\V)\times(\gl(\V)\oplus\V)\longrightarrow
\V$ is a bilinear map on the space of objects and on the space of morphisms. To see that it is a bilinear functor, we need
to prove that
\begin{enumerate}
\item[(a)] \label{itm:st} it preserves the source and target maps;
\item[(b)] \label{itm:id} it preserves  identity morphisms;
\item[(c)] \label{itm: comp} it preserves the composition of morphisms.
\end{enumerate}
Item $(\mathrm b)$ is obvious. Now we  give the proof of item
$(\mathrm a)$.

By (\ref{defi:pair1}), we have \begin{eqnarray*}
s\pair{A+\phi+u+m,B+\psi+v+n}&=&s\Big(\half\big((A+\phi\big)
(v+n)+\big(B+\psi) (u+m)\big)\Big)\\
&=&\half s \big((Av+An+\phi (v+\dM n)+Bu+Bm+\psi(u+\dM m)\big)\\
& =&\half (Av+Bu).
\end{eqnarray*}
By (\ref{defi:pair0}), we have
\begin{eqnarray*}
\pair{s(A+\phi+u+m),s(B+\psi+v+n)}&=&\pair{A+u,B+v}\\
&=&\half (Av+Bu)\\
& =&s\pair{A+\phi+u+m,B+\psi+v+n}.
\end{eqnarray*}
Thus the pairing $\pair{\cdot,\cdot}$ preserves the source map.

Similarly,  considering the target map, we have
 \begin{eqnarray*}
&&t\pair{A+\phi+u+m,B+\psi+v+n}\\&=&t\Big(\half\big((A+\phi\big)
(v+n)+\big(B+\psi) (u+m)\big)\Big)\\
&=&\half t \big((Av+An+\phi (v+\dM n)+Bu+Bm+\psi(u+\dM m)\big)\\
& =&\half (Av+Bu+\dM\circ An+\dM\circ \phi(v+\dM n)+\dM\circ
Bm+\dM\circ \psi(u+\dM m)),
\end{eqnarray*}
and
\begin{eqnarray*}
&&\pair{t(A+\phi+u+m),t(B+\psi+v+n)}\\
&=&\pair{A+\delta\phi+u+\dM m),B+\delta\psi+v+\dM n)}\\
&=&\half\big((A+\delta\phi)(v+\dM n)+(B+\delta\psi)(u+\dM m)\big)\\
&=&\half(Av+A\circ\dM n+\dM\circ\phi (v+\dM n)+Bu+B\circ\dM
m+\dM\circ\psi (u+\dM m)).
\end{eqnarray*}
 Since $A,B\in\End^0_\dM(\V)$ satisfy
\begin{equation}\label{eqn:composition}
\dM\circ A=A\circ \dM,\quad \dM\circ B=B\circ \dM,\end{equation} the
pairing $\pair{\cdot,\cdot}$  preserves the target map.

 At last, we
prove that the pairing also preserves the  composition of morphisms.
For any
$e_1=A+\phi+u+m,~e_1^\prime=A^\prime+\phi^\prime+u^\prime+m^\prime,~e_2=B+\psi+v+n,~e_2^\prime=B^\prime+\psi^\prime+v^\prime+
n^\prime$ satisfying
 \begin{eqnarray*}
t(e_1)=s(e_1^\prime),\quad t(e_2)=s(e_2^\prime).
\end{eqnarray*}
We have
\begin{eqnarray}
\label{eqn:temp1}A^\prime&=&A+\delta\phi,\quad u^\prime=u+\dM
m,\\\label{eqn:temp2} B^\prime&=&B+\delta\psi,\quad v^\prime=v+\dM
n,\\\nonumber e_1\cdot_\ve
e_1^\prime&=&A+\phi+\phi^\prime+u+m+m^\prime,
\end{eqnarray}
and
\begin{eqnarray*} e_2\cdot_\ve
e_2^\prime&=&B+\psi+\psi^\prime+v+n+n^\prime.
\end{eqnarray*}
Thus we have
\begin{eqnarray*}
\pair{e_1\cdot_\ve e_1^\prime,e_2\cdot_\ve
e_2^\prime}&=&\pair{A+\phi+\phi^\prime+u+m+m^\prime,
B+\psi+\psi^\prime+v+n+n^\prime}\\
&=&\half\big((A+\phi+\phi^\prime)(v+n+n^\prime)+
(B+\psi+\psi^\prime)(u+m+m^\prime) \big).
\end{eqnarray*}
On the other hand, we have
\begin{eqnarray*}
\pair{e_1,e_2}&=&\pair{A+\phi+u+m,
B+\psi+v+n+}\\
&=&\half\big((A+\phi)(v+n)+ (B+\psi)(u+m)
\big),
\end{eqnarray*}
and
\begin{eqnarray*}
\pair{e_1^\prime,e_2^\prime}&=&\pair{A^\prime+\phi^\prime+u^\prime+m^\prime,
B^\prime+\psi^\prime+v^\prime+n^\prime}\\
&=&\half\big((A^\prime+\phi^\prime)(v^\prime+n^\prime)+
(B^\prime+\psi^\prime)(u^\prime+m^\prime) \big).
\end{eqnarray*}
By straightforward computations, we have
\begin{eqnarray*}
t\pair{e_1,e_2}&=&\half t\big((A+\phi)(v+n)+ (B+\psi)(u+m) \big)\\
&=&\half\big(Av+\dM\circ An+\dM\circ\phi(v+\dM n)+Bu+\dM \circ
Bm+\dM\circ\psi(u+\dM m)\big),
\end{eqnarray*}
and
\begin{eqnarray*} s\pair{e_1^\prime,e_2^\prime}&=&\half
s\big((A^\prime+\phi^\prime)(v^\prime+n^\prime)+
(B^\prime+\psi^\prime)(u^\prime+m^\prime) \big)\\
&=&\half(A^\prime v^\prime+B^\prime u^\prime)\\
&=&\half\big((A+\delta\phi)(v+\dM n)+(B+\delta\phi)(u+\dM m)\big).
\end{eqnarray*}
By \eqref{eqn:composition}, since 
$\delta\phi=\dM\circ\phi$ on $V_0$, we have
$$
t\pair{e_1,e_2}=s\pair{e_1^\prime,e_2^\prime}.
$$
It is not hard to see that
\begin{eqnarray*}
\pair{e_1,e_2}\cdot_\ve\pair{e_1^\prime,e_2^\prime}&=&\half
\big((A+\phi)(v+n)+A^\prime
n^\prime+\phi^\prime(v^\prime+n^\prime)\\
&&+ (B+\psi)(u+m)+B^\prime m^\prime+\phi^\prime(u^\prime+m^\prime)
\big).
\end{eqnarray*}
By (\ref{eqn:temp1}), (\ref{eqn:temp2}), and the definition of
$\gl(\V)$ action \eqref{eqn:action},  we have
$$
A^\prime
n^\prime=(A+\delta\phi)(n^\prime)=(A+\phi\circ\dM)(n^\prime)=(A+\phi)(n^\prime),
$$
and $$ \phi^\prime(v^\prime+n^\prime)=\phi^\prime(v+\dM
n+n^\prime)=\phi^\prime(v+ n+n^\prime).
$$
After similar computations for $B^\prime
m^\prime+\phi^\prime(u^\prime+m^\prime)$, we obtain that
$$
\pair{e_1,e_2}\cdot_\ve\pair{e_1^\prime,e_2^\prime}=\pair{e_1\cdot_\ve
e_1^\prime,e_2\cdot_\ve e_2^\prime},
$$
which completes the proof. \qed\vspace{3mm}

Similar to Weinstein's definition of the bracket on omni-Lie algebras, we
introduce a skew-symmetric bilinear bracket operation, denote by
$\Courant{\cdot,\cdot}$, on $\gl(\V)\oplus\V$. On the space of
morphisms, it is given by
\begin{eqnarray}
\nonumber&&\Courant{A+\phi+u+m,B+\psi+v+n}\\\label{defi:bracket1}&=&
[A+\phi,B+\psi]+\half\big((A+\phi)(v+n)-(B+\psi)(u+m)\big).
\end{eqnarray}
On the space of objects, it is given by
\begin{equation}\label{defi:bracket0}
\Courant{A+u,B+v}=[A,B]+\half\big(Av-Bu\big).
\end{equation}

\begin{lem} The bracket operation
$\Courant{\cdot,\cdot}$ defined by \eqref{defi:bracket1} and
\eqref{defi:bracket0} is a bilinear functor.
\end{lem}
\pf It is straightforward to see that
 \begin{eqnarray*}
&&s\Courant{A+\phi+u+m,B+\psi+v+n}\\&=& s[A+\phi,B+\psi]+\half
s\big((A+\phi)(v+n)-(B+\psi)(u+m)\big)\\
&=&[A,B]+\half(Av-Bu).
\end{eqnarray*}
On the other hand, we have \begin{eqnarray*}
\Courant{s(A+\phi+u+m),s(B+\psi+v+n)}&=&\Courant{A+u,B+v}\\
&=&[A,B]+\half(Av-Bu).
\end{eqnarray*}
Therefore, the bracket operation $\Courant{\cdot,\cdot}$ preserves
the source map.

Now for the target map, we have
\begin{eqnarray*}
&&t\Courant{A+\phi+u+m,B+\psi+v+n}\\&=& t[A+\phi,B+\psi]+\half
t\big((A+\phi)(v+n)-(B+\psi)(u+m)\big)\\
&=&t[A+\phi,B+\psi]+\half(Av+A\circ\dM n+\dM\circ\phi(v+\dM
n)-Bu-\dM\circ Bm-\dM\psi(u+\dM m)),\end{eqnarray*} and
\begin{eqnarray*}
&&\Courant{t(A+\phi+u+m),t(B+\psi+v+n)}\\
&=&\Courant{t(A+\phi)+u+\dM m,t(B+\psi)+v+\dM n}\\
&=&[t(A+\phi),t(B+\psi)]+\half\big((A+\delta\phi)(v+\dM
n)-(B+\delta\psi)(u+\dM m)\big).
\end{eqnarray*}
Since the bracket operation $[\cdot,\cdot]$ on $\gl(\V)$ is a
bilinear functor, we have
$$
[t(A+\phi),t(B+\psi)]=t[A+\phi,B+\psi].
$$
By (\ref{eqn:composition}), we see that the bracket operation
$\Courant{\cdot,\cdot}$ also preserves the target
 map.

It is obvious that $\Courant{\cdot,\cdot}$ preserves the identity morphism. Similar to the proof
 of Lemma \ref{lem:pair}, one can show that $\Courant{\cdot,\cdot}$
 also preserves  the composition of morphisms.
 Thus the bracket operation $\Courant{\cdot,\cdot}$ is a bilinear functor. \qed\vspace{3mm}

Since the nondegenerate symmetric pairing $\pair{\cdot,\cdot}$ and
the bracket operation $\Courant{\cdot,\cdot}$ are bilinear functors,
they are totally determined by the values on the space of morphisms,
i.e. they are determined by (\ref{defi:pair1}) and
(\ref{defi:bracket1}).

\begin{defi}\label{defi:omni2}
The triple
$(\gl(\V)\oplus\V,\pair{\cdot,\cdot},\Courant{\cdot,\cdot})$ is
called the omni-Lie $2$-algebra associated to the $2$-vector space
$\V$, where $\pair{\cdot,\cdot}$ is the symmetric bilinear functor
given by \eqref{defi:pair1} and $\Courant{\cdot,\cdot}$ is the
skew-symmetric bilinear functor given by \eqref{defi:bracket1}. We
simply denote the omni-Lie $2$-algebra by $\gl(\V)\oplus\V$.
\end{defi}

 The factor of $\half$ in
$\Courant{\cdot,\cdot}$ spoils the Jacobi identity. Computing the
Jacobi identity of the bracket operation $\Courant{\cdot,\cdot}$ on
the space of objects, we have
\begin{eqnarray*}
\Courant{\Courant{A+u,B+v},C+w}+c.p.&=&\frac{1}{4}\big([A,B]w+[B,C]u+[C,A]v\big).
\end{eqnarray*}
Thus in general the omni-Lie $2$-algebra
$(\gl(\V)\oplus\V,\pair{\cdot,\cdot},\Courant{\cdot,\cdot})$ is not
a Lie $2$-algebra, since the Jacobiator identity is not
satisfied.\vspace{3mm}

We can also introduce another bracket operation
$\Dorfman{\cdot,\cdot}$ on $\gl(\V)\oplus\V$, which is not
skew-symmetric, by setting
\begin{equation}\label{relation}
   \Dorfman{e_1,e_2}\triangleq\Courant{e_1,e_2}+\pair{e_1,e_2}.
\end{equation}
Since $\Courant{\cdot,\cdot}$ and $\pair{\cdot,\cdot}$ are all
bilinear functors, $\Dorfman{\cdot,\cdot}$ is also a bilinear
functor. Assume $e_1=A+\phi+u+m$ and $e_2=B+\psi+v+n$, we have
\begin{eqnarray}\label{def:bracketdorfman}
\label{defi:lbracket1}\Dorfman{A+\phi+u+m,B+\psi+v+n}&=&
[A+\phi,B+\psi]+(A+\phi)(v+n).
\end{eqnarray}

By straightforward computations, we have
\begin{pro}\label{pro:leibniz}
The bracket operation $\Dorfman{\cdot,\cdot}$ satisfies
the Leibniz rule, i.e. for any $e_1,e_2,e_3\in\gl(\V)\oplus \V$, we
have
\begin{eqnarray*}
\Dorfman{e_1,\Dorfman{e_2,e_3}} =\Dorfman{\Dorfman{e_1,e_2},e_3}
+\Dorfman{e_2,\Dorfman{e_1,e_3}}.
\end{eqnarray*}
\end{pro}

\begin{rmk}
  In \cite{shengliuTCA}, the first two authors introduce the notion of  Leibniz
  $2$-algebras, which is equivalent to $2$-term $Leibniz_\infty$-algebras \cite{UchinoshLeibniz}. In fact, Proposition \ref{pro:leibniz} implies that $(\gl(\V)\oplus\V,
  \Dorfman{\cdot,\cdot})$ is a strict Leibniz $2$-algebras, please see \cite{shengliuTCA} for more details.
\end{rmk}

For a 2-sub-vector space $\LL\subset\gl(\V)\oplus\V$, define
$\LL^\perp$ by
\begin{equation}
\LL^\perp=\{e\in\gl(\V)\oplus\V|~\pair{e,l}=0,~\forall~l\in\LL\}.
\end{equation}

 Dirac
structures of the omni-Lie $2$-algebra
$(\gl(\V)\oplus\V,\pair{\cdot,\cdot},\Courant{\cdot,\cdot})$ are
defined in the usual way.

\begin{defi}
A Dirac structure of the  omni-Lie $2$-algebra
$(\gl(\V)\oplus\V,\pair{\cdot,\cdot},\Courant{\cdot,\cdot})$ is a
maximal isotropic $2$-sub-vector space, i.e. $\LL=\LL^\perp$, which
is closed under the bracket operation $\Courant{\cdot,\cdot}$.
\end{defi}

\begin{pro}\label{pro:Dirac2algebra}
Let $D$ be a Dirac structure of the  omni-Lie $2$-algebra
$\gl(\V)\oplus\V$. Then $(D, \Courant{\cdot,\cdot})$ is a strict Lie
$2$-algebra.
\end{pro}
\pf By (\ref{relation}), a maximal isotropic $2$-sub-vector space is
closed under  $\Courant{\cdot,\cdot}$ if and only if it is closed
under $\{\cdot,\cdot\}$.
 By \eqref{relation} and  Proposition
\ref{pro:leibniz}, $(D,\Courant{\cdot,\cdot})=(D,\{\cdot,\cdot\})$
is a strict Lie $2$-algebra.\qed\vspace{3mm}

For any bilinear functor $F:\V\times \V\longrightarrow\V$, define
$\ad_F:\V\longrightarrow\gl(\V)$ by
$$\ad_F\xi(\eta)=F(\xi,\eta),\quad \forall~\xi,\eta\in\V.$$


\begin{lem}
For any bilinear functor $F:\V\times \V\longrightarrow\V$, the induced map $\ad_F:\V\longrightarrow\gl(\V)$ is a linear functor.
\end{lem}
\pf Since $F$ is a bilinear functor, it is straightforward to see
that
$$
\ad_Fu\in\End^0_\dM(\V),\quad \ad_Fm\in \End^1(\V), \quad\forall~
u\in V_0,~m\in V_1,
$$
which implies that $\ad_F$ preserves the identity morphisms.
Furthermore, we have
$$
s(\ad_F(u+m))=\ad_Fu=\ad_F(s(u+m)),
$$
which implies that $\ad_F$ preserves the source map. For the target
map, we have
\begin{eqnarray*}
t(\ad_F(u+m))&=&\ad_Fu+\delta(\ad_Fm),\\
\ad_F(t(u+m))&=&\ad_Fu+\ad_F\dM m,\\
\delta(\ad_Fm)(v)&=&\dM (\ad_Fm(v))= \dM F(m,v), \quad \forall v \in
V_0, \\
\delta(\ad_Fm)(n)&=& \ad_Fm (\dM n)=F(m,\dM n), \quad \forall n \in
V_1.
\end{eqnarray*}
According to Theorem 4.3.6 in \cite{baez:2algebras} and Proposition
2.6 in \cite{roytenbergweakLie2}, for any bilinear functor
$F:\V\times \V\longrightarrow\V$, we have
$$
\dM F(m,v)=  F(\dM m,v),\quad F(m,\dM n)=F(\dM m, n).
$$
Therefore, we have
\begin{equation}\label{eqn:deltad}
\delta(\ad_Fm)=\ad_F\dM m,
\end{equation}
which implies that $\ad_F$ preserves the target map.

Finally, we prove that $\ad_F$ also preserves the composition of
morphisms. For any morphisms $u+m$ and $v+n$ satisfying
$t(u+m)=s(v+n)$, i.e $v=u+\dM m$, we have
$$
(u+m)\cdot_\ve(v+n)=u+m+n.
$$
It is straightforward to see that
\begin{eqnarray*}
\ad_F((u+m)\cdot_\ve(v+n))&=&\ad_F(u+m+n)=\ad_Fu+\ad_F(m+n).
\end{eqnarray*}
By (\ref{eqn:deltad}), we have
\begin{eqnarray*}
t\ad_F(u+m)&=&\ad_Fu+\delta\ad_Fm=\ad_Fu+\ad_F\dM m=\ad_F(u+\dM
m).
\end{eqnarray*}
Obviously we have
\begin{eqnarray*}
 s\ad_F(v+n)&=&\ad_Fv=\ad_F(u+\dM m).
\end{eqnarray*}
Thus we have $s\ad_F(v+n)=t\ad_F(u+m)$ and
$$
\ad_F(u+m)\cdot_\ve\ad_F(v+n)=\ad_Fu+\ad_Fm+\ad_Fn=\ad_Fu+\ad_F(m+n),
$$
which yields that $\ad_F$ preserves the composition of morphisms.
 \qed\vspace{3mm}

 We denote by $\frkG_F\subset\gl(\V)\oplus\V$ the graph of the
operator $\ad_F$.

\begin{thm}\label{Thm:graphdirac}
Given a bilinear functor $F:\V\times \V\longrightarrow\V$, the graph $\frkG_F$
is a Dirac structure of the
 omni-Lie $2$-algebra $\gl(\V)\oplus\V$ if and only if $F$ defines a strict Lie $2$-algebra structure on the $2$-vector
space $\V$.
\end{thm}
\pf For any $\xi,\eta\in\V,$ we have
$$\pair{\ad_F\xi+\xi,\ad_F\eta+\eta}=\half\big(\ad_F\xi(\eta)+\ad_F\eta(\xi)\big)=\half\big(F(\xi,\eta)+F(\eta,\xi)\big).$$
Thus $\frkG_F$ is isotropic iff $F$ is skew-symmetric. Take any $(A,
\eta) \in \frkG_F^\perp$, if $F$ is skew-symmetric, then
\[ 0=2\pair{\ad_F \xi + \xi ,  A +\eta}=F(\xi, \eta) + A(\xi) =- \ad_F
\eta (\xi) +A(\xi).
\]This implies that $A= \ad_F \eta$, thus $(A, \eta) \in
\frkG_F$. Hence $\frkG_F$ is actually maximal isotropic. By this
argument, we see that $\frkG_F$ is maximal isotropic iff $F$ is
skew-symmetric.

Moreover, for any skew-symmetric bilinear functor $F$, we have
\begin{eqnarray*}
\Courant{\ad_F\xi+\xi,\ad_F\eta+\eta}&=&[\ad_F\xi,\ad_F\eta]+\half\big(\ad_F\xi(\eta)-\ad_F\eta(\xi)\big)\\
&=&[\ad_F\xi,\ad_F\eta]+F(\xi,\eta).
\end{eqnarray*}
Thus $\frkG_F$ being a Dirac structure is equivalent to the fact that $F$ is
skew-symmetric and
satisfies \begin{equation}\label{eq:imply-jacobi}
[\ad_F\xi,\ad_F\eta]=\ad_FF(\xi,\eta).
\end{equation}
 These two
properties are also equivalent to the fact that $F$ provides a
strict Lie $2$-algebra structure on $\V$: indeed, applying
\eqref{eq:imply-jacobi} to an element $\theta \in \V$ gives the
Jacobi identity
$$
F(\xi,F(\eta,\theta))=F(F(\xi,\eta),\theta)+F(\eta,F(\xi,\theta)),
$$
provided that $F$ is skew-symmetric. \qed\vspace{3mm}

In \cite{clomni}, Chen and Liu introduce the notion of omni-Lie
algebroid associated to a vector bundle $E$,  which generalizes
Weinstein's omni-Lie algebras associated a vector space $V$. They
continue to study Dirac structures of omni-Lie algebroids in
\cite{clsdirac}. In that paper, they show that  there is a
one-to-one correspondence between Dirac structures (not necessarily
coming from graphs) and projective Lie algebroids. This implies
that, in particular, for the omni-Lie algebra $\gl(V)\oplus V$,
there is a one-to-one correspondence between Dirac structures (not
necessarily coming from graphs) and Lie algebra structures on the
sub-vector space of $V$.\vspace{3mm}

For omni-Lie $2$-algebras, we have

\begin{thm}\label{thm:general Dirac}
There is a one-to-one correspondence between Dirac structures of the
omni-Lie $2$-algebra
$(\gl(\V)\oplus\V,\pair{\cdot,\cdot},\Courant{\cdot,\cdot})$ and
strict Lie $2$-algebra structures on $2$-sub-vector spaces of $\V$.
\end{thm}
To prove this theorem, we need to adapt the theory of
characteristic pairs developed in \cite{liuDirac} for Courant
algebroids to our setting. This theory has  many applications in the theory of
reduction of various geometric structures, see for example \cite{reductionJacobi}.

Given a maximal isotropic $2$-sub-vector space $\LL$ of the omni-Lie
$2$-algebra
$(\gl(\V)\oplus\V,\pair{\cdot,\cdot},\Courant{\cdot,\cdot})$, let
$\D=\LL\cap\gl(\V)$. Obviously, $\D$ is a $2$-sub-vector space. We
define the $2$-sub-vector space $\D^0\subset\V$ to be the null space
of $\D$:
$$
\D^0=\{\xi\in\V~|\quad X(\xi)=0,\quad\forall~X\in \D\}.
$$
Similarly, for any 2-sub-vector space $\W\subset \V$, we define
$\W^0$ by
$$
\W^0=\{D\in\gl(\V)~|\quad D(\xi)=0,\quad\forall~\xi\in \W\}.
$$
It is straightforward to  see that

\begin{lem}\label{lem:0}
  With the above notations, we have
  $$
D\subset(D^0)^0,\quad (\W^0)^0=\W.
  $$
\end{lem}

 If $\LL$ is maximal isotropic, then $\LL$ is of the form
$$
\LL=\D\oplus
\frkG_{\pi|_{\D^0}}=\{X+\pi(\xi)+\xi~|X\in\D,~\xi\in\D^0\},
$$
for some linear functor $\pi:\V\longrightarrow \gl(\V)$ satisfying
\begin{equation}\label{eq:pi-xi-eta}
\pi(\xi)(\eta)=-\pi(\eta)(\xi).
\end{equation}

Let us explain this: $\D$ is the kernel of the projection
$\pr_\V:\gl(\V)\oplus\V\longrightarrow\V$ restricted to $\LL$. Take
any splitting $\LL=\D\oplus\D^\prime$. Denote by $\HH$ the image of
$\pr_\V|_\LL$. Then $\pr_\V:{\D^\prime}\longrightarrow \HH$ is
bijective, thus $\D^\prime$ is the graph of some linear functor,
$\pi:\V\longrightarrow\gl(\V)$, restricted to $\HH$. Therefore, we
have
$$ \LL=\D\oplus
\frkG_{\pi|_{\HH}}.
$$
For any $Y\in \gl(\V)$, since $\pair{X+\pi(\xi)+\xi, Y}=\half
Y(\xi)$, it is easy to see that $Y$ is in $\HH^0$ iff $Y$ is in
$\LL^\perp$. Since $\LL$ is maximal isotropic, we have
$\LL^\perp=\LL$. Thus $\HH^0 = \D$, which implies that $\HH=\D^0$.
For any $\eta, \xi \in\D^0$, since $\LL$ is isotropic, we have
$$
\pair{\pi(\xi)+\xi,\pi(\eta)+\eta}=\half\big(\pi(\xi)(\eta)+\pi(\eta)(\xi)\big)=0,
$$
which implies that $\pi(\xi)(\eta)=-\pi(\eta)(\xi)$.

Clearly the function $\pi$ depends on the choice of the splitting
$\LL = \D \oplus \D^\prime$. Such a pair $(\D,\pi)$ is called a {\bf
characteristic pair} of the maximal isotropic $2$-sub-vector space
$\LL$. Two characteristic pairs $(\D_1,\pi_1)$ and $(\D_2,\pi_2)$
determine the same maximal isotropic $2$-sub-vector space $\LL$ iff
$$
\D_1=\D_2, \quad {\pi_1}(\xi)-{\pi_2}(\xi)\in \D,\quad\forall~\xi\in
\D^0.
$$

The conditions under which $\LL=\D\oplus \frkG_{\pi|_{\D^0}}$ is a
Dirac structure is given by the following proposition.

\begin{pro}\label{pro:condition of Dirac}
Let $(\D, \pi)$ be a characteristic pair of a maximal isotropic
$2$-sub-vector space $\LL$ of $\gl(\V)\oplus \V$. Then $\LL=\D\oplus
\frkG_{\pi|_{\D^0}}$ is a Dirac structure if and only if for any
$~\xi,\eta \in\D^0$, the following conditions are satisfied:
\begin{itemize}
\item[\rm(1)]~$ \D$ is a sub-Lie  $2$-algebra of $\gl(\V)$;
\item[\rm(2)]~$\pi\big(\pi(\xi)(\eta)\big)-[\pi(\xi),\pi(\eta)]\in \D$;
\item[\rm(3)]~$\pi(\xi)(\eta)\in\D^0$.
\end{itemize}
\end{pro}
\pf Suppose that $\LL$ is a Dirac structure. For any  $X,Y\in\D$, we
have $ \Courant{X,Y}=[X,Y]. $ Thus  $\LL$ being a Dirac structure
implies that $[X,Y]\in\LL\cap\gl(\V)=\D$, so $\D$ is a sub-Lie
$2$-algebra of $\gl(\V)$. For any $X, Y \in \D, ~\eta\in\D^0$, we
have
$$
\Courant{X,Y+\pi(\eta)+\eta}=[X,Y]+[X,\pi(\eta)]+\half
X(\eta)=[X,Y]+[X,\pi(\eta)] \in \LL,
$$
which implies that $[X,\pi(\eta)]\in\LL\cap\gl(\V)=\D$. Similarly,
for any $\xi, \eta \in \D^0$, we have
$$
\Courant{\pi(\xi)+\xi,\pi(\eta)+\eta}=[\pi(\xi),\pi(\eta)]+\pi(\xi)(\eta)
\in \LL,
$$which implies that $\pi(\xi)(\eta) \in \D^0$ and
$[\pi(\xi),\pi(\eta)]-\pi\big(\pi(\xi)(\eta)\big) \in \D$.

Conversely, for any $X,Y\in\D,~\xi,\eta \in\D^0$, we have
$$
\Courant{X+\pi(\xi)+\xi,Y+\pi(\eta)+\eta}=[X,Y]+[X,\pi(\eta)]+[\pi(\xi),Y]+[\pi(\xi),\pi(\eta)]+\pi(\xi)(\eta).
$$
Then for any $\theta \in \D^0$, $[\pi(\xi), Y] (\theta)= \pi(\xi)
(Y(\theta)) - Y(\pi(\xi)(\theta)) =0$ by Condition $(3)$. Thus $
[\pi(\xi), Y]\in \D$. Then it is straightforward to see that $\LL$
is a Dirac structure if Conditions $(1),(2),(3)$ are satisfied. This
concludes the proof. \qed\vspace{3mm}

{\bf The proof of Theorem \ref{thm:general Dirac}:} For any Dirac
structure $\LL=\D\oplus\frkG_{\pi|_{\D^0}}$, define a bilinear
functor, $[\cdot,\cdot]_{\D^0}:\D^0\times\D^0\longrightarrow\D^0$ on
$\D^0$ by
$$
[\xi,\eta]_{\D^0}\triangleq\pi(\xi)(\eta),\quad \forall~\xi,\eta\in
\D^0.
$$
By Condition (3) in Proposition \ref{pro:condition of Dirac}, it is
well defined, and by \eqref{eq:pi-xi-eta} it is skew-symmetric. By
Condition (2), we have for all $\xi, \eta, \theta \in \D^0$,
 \begin{eqnarray*}
 [[\xi,\eta]_{\D^0},\theta]_{\D^0}&=&\pi([\xi,\eta]_{\D^0})(\theta)=\pi\big((\pi(\xi)(\eta)\big)(\theta)=[\pi(\xi),\pi(\eta)](\theta)\\&=&\pi(\xi)(\pi(\eta)(\theta))
 -\pi(\eta)(\pi(\xi)(\theta))\\
 &=&[\xi,[\eta,\theta]_{\D^0}]_{\D^0}-[\eta,[\xi,\theta]_{\D^0}]_{\D^0},
\end{eqnarray*}
which implies that $(\D^0,[\cdot,\cdot]_{\D^0})$ is a strict Lie
$2$-algebra. Thus any Dirac structure gives rise to a strict Lie
$2$-algebra structure on a $2$-sub-vector space of $\V$.

Conversely, for any $2$-sub-vector space $\W$ of $\V$, assume that
$\V=\W\oplus\W^\prime$. Define $\D$ by
$$
\D=\W^0\triangleq\{X\in\gl(\V)|~X(\xi)=0,\quad\forall~\xi\in\W\}.
$$ Then by Lemma \ref{lem:0}, we have $\D^0 =\W$.
Obviously, for any $D,D^\prime\in \D$ and $\xi\in\W$, we have
$$
[D,D^\prime](\xi)=0,
$$
which implies that $\D$ is a sub-Lie $2$-algebra of $\gl(\V)$. By
the inclusion $\W \to \V$, we have a natural embedding $\gl(\W)
\subset \gl(\V)$ as a sub-Lie $2$-algebra. For any Lie $2$-algebra
structure $[\cdot,\cdot]_\W$ on $\W$, we have a linear functor
$\ad:\W\longrightarrow\gl(\W)$ which is given by
$\ad_\xi(\eta)=[\xi,\eta]_\W$. Define
$\huaF:\V\longrightarrow\gl(\V)$, as an extension of $\ad$, by
setting
$$
\huaF(\xi+\xi^\prime)\triangleq\ad_\xi,\quad\forall~\xi\in\W,~\xi^\prime\in\W^\prime.
$$
Now we consider the $2$-sub-vector space
$\LL\triangleq\D\oplus\frkG_{\huaF|_\W}$, which is the direct sum of
$\D$ and the graph of $\huaF|_\W$.
 Since $[\cdot,\cdot]_\W$ is skew-symmetric, $\LL$ is
isotropic. Take $A+\eta \in \LL^\perp$. First by $\pair{X,A+\eta}=0$
for all $X\in \D$, we have $\eta\in \D^0$. Thus for all $X\in \D,
\xi \in \D^0$, we have
$$
\pair{X+\huaF(\xi)+\xi,A+\eta}=\huaF(\xi)(\eta)+A\xi=-\huaF(\eta)(\xi)+A\xi=0,
$$
which implies that $A=\huaF(\eta)+Y$, for some $Y\in \D$. Thus $A+\eta\in \D+\frkG_\huaF|_\W$.  By the fact that $[\cdot,\cdot]_\W$ satisfies the Jacobi
identity, we obtain
$$[\ad_\xi,\ad_\eta]=\ad_{[\xi,\eta]_\W}=\ad_{\ad_\xi \eta} ,\quad\forall~\xi,\eta\in\W.$$
Furthermore, for any $D\in \D$ and $\xi,\eta\in\W$, it is obvious
that
$$
[D,\ad_\xi](\eta)=D([\xi,\eta]_\W)-[\xi,D(\eta)]=0,
$$
which yields that $[D,\ad_\xi]\in\D$. Therefore, we have
\begin{eqnarray*}
\Courant{D+\ad_\xi+\xi,D^\prime+\ad_\eta+\eta}&=&[D,D^\prime]+[D,\ad_\eta]+[\ad_\xi,D^\prime]+[\ad_\xi,\ad_\eta]+\half(\ad_\xi(\eta)-\ad_\eta(\xi))\\
&=&[D,D^\prime]+[D,\ad_\eta]+[\ad_\xi,D^\prime]+\ad_{[\xi,\eta]_\W}+[\xi,\eta]_\W\\
&\in&\D\oplus\frkG_{\huaF|_\W},
\end{eqnarray*}
which yields that $\LL$ is closed under the bracket operation
(\ref{defi:bracket1}). Thus $\LL$ is a Dirac structure. Obviously a different extension $\huaF'$ of $\ad$ gives rise to the same Dirac structure.
 \qed

\section{Normalizer of a Dirac structure}

In this section, we introduce the notion of the normalizer of a
$2$-sub-vector space of the omni-Lie $2$-algebra $\gl(\V)\oplus \V$.
In the classical case, the normalizer of the graph of the adjoint
operator of a Lie algebra $\frkg$ is the derivation Lie algebra
$\Der(\frkg)$. Here we will prove  a similar result for strict Lie
$2$-algebras.

\begin{defi}
The normalizer of a  $2$-sub-vector space $\K\subset\gl(\V)\oplus
\V$ is composed of all the elements $N\in \gl(\V)$ such that
$$\Dorfman{N,\K}\subset\K,$$
with $\Dorfman{\cdot, \cdot}$ defined in \eqref{relation}.
\end{defi}

Denote by $N_\K$ the normalizer of $\K$. Especially we care about
the normalizer of a Dirac structure $\LL$.

\begin{pro}\label{pro:normalizer}
Let $\LL=\D \oplus \frkG_{\pi|_{\D^0}}$ be a Dirac structure of the
omni-Lie $2$-algebra $\gl(\V) \oplus \V$ with characteristic pair
$(\D, \pi)$. Then by Theorem \ref{thm:general Dirac}, $\D^0$ is a
strict Lie $2$-algebra with Lie bracket
$[\xi,\eta]_{\D^0}=\pi(\xi)(\eta) $.
 Now we claim that $N \in N_\LL$ if and only if
\begin{enumerate}
 \item for all $X \in \D$, we have $[N, X]\in \D$,
\item for all $\xi, \eta \in \D^0\subset \V$, we have $N(\xi) \in \D^0$ and $N([\xi, \eta]_{\D^0}) =[\xi, N(\eta)]_{\D^0} + [N(\xi), \eta]_{\D^0} $.
\end{enumerate}
\end{pro}
\begin{proof}
By definition, $N \in N_\LL$ if and only if for all $X\in \D$ and
$\xi \in \D^0$, we have \[\{ N, X +\pi(\xi) +\xi \} = [N,X] + [N,
\pi(\xi)] + N(\xi) \in \LL ,\] which is equivalent to the fact that
$[N, X] \in \D$ for all $X \in \D$ (by taking $\xi=0$), and $N(\xi)
\in \D^0$ and $\pi(N(\xi))-[N, \pi(\xi)] \in \D$ for all $\xi \in
\D^0$.  The fact that $\pi(N(\xi))-[N, \pi(\xi)] \in \D$ for all
$\xi \in \D^0$, is equivalent to the fact that, for all $\eta, \xi
\in \D^0$,
\[
0=\pi(N(\xi))(\eta)-[N, \pi(\xi)] (\eta) = [N(\xi), \eta]_{\D^0} -
N([\xi, \eta]_{\D^0}) + [\xi, N(\eta)]_{\D^0}.
\]
This finishes the proof.
\end{proof}

It is  subtle to define  sub-Lie $2$-algebras of a Lie $2$-algebra.
At first sight,  we might define a sub-Lie $2$-algebra of a Lie
$2$-algebra to be a $2$-sub-vector space which is closed under the
bracket operation (Baez et al. use this definition in
\cite{baez:string}). Then one can also propose that a sub-Lie
$2$-algebra $L'$ of a Lie $2$-algebra $L$ is an injective morphism
$\mu : L' \to L$ (see Definition \ref{def:morphism}). These two
definitions are not the same. The second definition gives the first
definition iff $\mu$ is strict. We clarify here that we use the
first definition  in this paper, even though in \cite{shengzhu2} we
use the second more general definition.\vspace{3mm}

\begin{pro}\label{pro:normlizer-subl2}
The normalizer $N_\LL$ of a Dirac structure $\LL$ of an omni-Lie
$2$-algebra $\gl(\V)\oplus \V$ associated to a $2$-vector space $\V$
is a sub-Lie $2$-algebra of $\gl(\V)$.
\end{pro}
\begin{proof}
First of all we show that $N_\LL$ is a sub-$2$-vector space of
$\gl(\V)$. For this, we only need to verify that $\delta(N_\LL \cap
\End^1(\V)) \subset N_\LL $.

Take $\phi \in N_\LL \cap \End^1(\V)  $. By Proposition \ref{pro:normalizer}, and using the same notation therein,  for all $X\in \D$, we have
\[
[\delta (\phi), X] = \delta([\phi, X])-[\phi, \delta(X)] \in \D,
 \] since $(\End^1(\V)\xrightarrow{\delta} \End_\dM^0 (\V), [\cdot,\cdot ])$ is a DGLA and $\D$ is a $2$-sub-vector space of $\gl(\V)\oplus \V$.  Then for all $X\in \D$,
  $\xi\in \D^0$, we have
\[
 X(\delta(\phi)(\xi))=[X, \delta (\phi)](\xi)+\delta(\phi)(X(\xi))=[X, \delta(\phi)](\xi) = 0.
\] Thus $\delta(\phi)(\xi) \in \dperp$. Now for all  $u, v\in \D^0\cap V_0$, $n \in \D^0\cap V_1$,  we have
\begin{eqnarray*}
\delta(\phi) ([u,v]_{\D^0}) &=&\dM\circ\phi([u,v]_{\D^0})=\dM([\phi(u),v]_{\D^0})+\dM([u,\phi(v)]_{\D^0})\\
&=&[\dM(\phi(u)), v]_{\D^0}+[u,\dM( \phi(v))]_{\D^0}\\
&=&[\delta(\phi)(u),v]_{\D^0}+[u,\delta(\phi)(v)]_{\D^0},
 \end{eqnarray*}
 and
  \begin{eqnarray*}
 \delta (\phi)([u,n]_\dperp)&=&\phi\circ\dM ([u,n]_\dperp)=\phi [\dM u, n]_\dperp\\
 &=& [\phi(\dM u), n]_\dperp+[\dM u,\phi( n)]_\dperp\\
 &=& [\delta(\phi)(u),n]_\dperp+ [u,,\delta(\phi)(n)]_\dperp.
 \end{eqnarray*}
Thus by Proposition \ref{pro:normalizer}, $\delta(\phi) \in N_\LL$.

To see further  that $N_\LL$ is a sub-Lie $2$-algebra, we only need
to prove that $N_\LL$ is closed under the induced bracket operation.
For all $N,N^\prime\in N_\LL$, $l \in \LL$, we have
\begin{eqnarray*}
  \Dorfman{\Dorfman{N,N^\prime},l}=\Dorfman{[N,N^\prime],l}.
\end{eqnarray*}
On the other hand, by Proposition \ref{pro:leibniz}, we have
\begin{eqnarray*}
  \Dorfman{\Dorfman{N,N^\prime},l}=\Dorfman{N,\Dorfman{N^\prime,l}}-\Dorfman{N^\prime,\Dorfman{N,l}}\in \LL.
\end{eqnarray*}
Therefore, $[N,N^\prime]\in N_\LL$.
\end{proof}

As a corollary of the above two propositions, we have,

\begin{cor}\label{thm:normalizer}
Given a strict Lie $2$-algebra $\V$,  the normalizer $N_{\frkG_\ad}$
of the Dirac structure  $\frkG_\ad$ is a sub-Lie $2$-algebra of
$\gl(\V)$, and  $D\in N_{\frkG_\ad}$ if and only if for any
$\xi,\eta\in\V$, we have
\begin{equation}\label{eqn:derivation}
D[\xi,\eta]=[D\xi,\eta]+[\xi,D\eta].
\end{equation}
\end{cor}

\begin{rmk}
In some unpublished works\footnote{Private conversation with
Stevenson.},  Stevenson and Schlessinger-Stasheff study the notion
of  derivations of  Lie $n$-algebras. Here by studying the
normalizer of the graph of adjoint operator, we recover their notion
of derivation: the Lie $2$-algebra $N_{\frkG_\ad}$ is exactly the
Lie $2$-algebra of derivations $\Der(\V)$ for a Lie $2$-algebra
$\V$.
\end{rmk}

\section{Twisted omni-Lie $2$-algebras}

In this section we introduce the notion of twisted  omni-Lie
$2$-algebra $\gl(\V)\oplus_\mu\V$, where
$\mu:\gl(\V)\longrightarrow\gl(\V)$ is an isomorphism of Lie
$2$-algebras. We will show that Dirac structures of the twisted
omni-Lie $2$-algebra $\gl(\V)\oplus_\mu\V$ characterize some special
Lie $2$-algebra structures on $\V$.

\begin{defi} \label{def:morphism} {\em\cite{baez:2algebras}}
Given Lie $2$-algebras $C$ and $C^\prime$, a Lie $2$-algebra
morphism $\mu:C\longrightarrow C^\prime$ consists of:
\begin{itemize}
\item[$\bullet$] a linear functor $(\mu_0,\mu_1)$ from the underlying $2$-vector space
of $C$ to that of $C^\prime$, and
\item[$\bullet$] a skew-symmetric bilinear natural transformation
$$
\mu_2(u,v):\mu_0[u,v]\longrightarrow [\mu_0(u),\mu_0(v)]
$$
\end{itemize}
such that the following diagram commutes: \begin{equation}\label{eq:higher-constraints-mu2}
\xymatrix{
\mu_0[[u,v],w]\ar[d]_{\mu_2([u,v],w)}\ar[rr]^{\mu_0J_{u,v,w}\quad\qquad\quad}&&\mu_0[u,[v,w]]+\mu_0[v,[w,u]]\ar[d]^{\mu_2(u,[v,w])+\mu_2(v,[w,u])}\\
~[\mu_0[u,v],\mu_0(w)]\ar[d]_{[\mu_2(u,v),id_{\mu_0(w)}]}&&[\mu_0(u),\mu_0[v,w]]+[\mu_0(v),\mu_0[w,u]]\ar[d]^{[id_{\mu_0(u)},\mu_2(v,w)]+[id_{\mu_0(v)},\mu_2(w,u)]}\\
~[[\mu_0(u),\mu_0(v)],\mu_0(w)]\ar[rr]^{J\qquad\quad}&&
[\mu_0(u),[\mu_0(v), \mu_0(w)]]+[\mu_0(v),[\mu_0(w),\mu_0(u)]]. }
\end{equation}
We call $\mu$ \underline{strict} if $\mu_2=0$. We call $\mu$ an
\underline{isomorphism} if  it induces an isomorphism of the
underlying $2$-vector spaces.
\end{defi}

Now we take an isomorphism  $\mu$ from the Lie $2$-algebra $\gl(\V)$
to itself. We define the following $\mu$-twisted bracket
$\Courant{\cdot,\cdot}_\mu$ on $\gl(\V)\oplus \V$:
\begin{eqnarray}\label{defi:twistb}
\nonumber&&\Courant{A+\phi+u+m,B+\psi+v+n}_\mu\\&&=
[A+\phi,B+\psi]+\half\big(\mu_1(A+\phi)(v+n)-\mu_1(B+\psi)(u+m)\big).
\end{eqnarray}
The nondegenerate $\V$-valued pairing can also be twisted by $\mu$,
\begin{eqnarray}\label{defi:twistp}
\pair{A+\phi+u+m,B+\psi+v+n}_\mu=
\half\big(\mu_1(A+\phi)(v+n)+\mu_1(B+\psi)(u+m)\big).
\end{eqnarray}

\begin{defi}\label{defi:omniLie2twist}
The triple
$(\gl(\V)\oplus\V,\Courant{\cdot,\cdot}_\mu,\pair{\cdot,\cdot}_\mu)$
is called a $\mu$-twisted omni-Lie $2$-algebra. We simply denote it
by $\gl(\V)\oplus_\mu\V$.
\end{defi}We also introduce the bracket $\Dorfman{\cdot,\cdot}_\mu$ by
setting
\begin{eqnarray}
\label{dorfmanmu}
\Dorfman{\cdot,\cdot}_\mu=\Courant{\cdot,\cdot}_\mu+\pair{\cdot,\cdot}_\mu.
\end{eqnarray}

It is not hard to see that
$\Courant{\cdot,\cdot}_\mu~\pair{\cdot,\cdot}_\mu$ and
$~\Dorfman{\cdot,\cdot}_\mu$ are all bilinear functors since these
operations without $\mu$-twist are bilinear functors and $\mu$ is an
isomorphism of the strict Lie $2$-algebra $\gl(\V)$.

\begin{pro}\label{pro:jacobiator}
We have a natural transformation $J$ between the functors
$\{\{\cdot,\cdot \}_\mu, \cdot\}_\mu$ and $\{\cdot, \{\cdot,
\cdot\}_\mu \}_\mu-\{\cdot, \{\cdot,\cdot \}_\mu \}_\mu$ defined as
follows: For any objects $e_1=A+u,~e_2=B+v,~e_3=C+w$ in
$\gl(\V)\oplus\V$,
$$J_{e_1,e_2,e_3}:\Dorfman{\Dorfman{e_1,e_2}_\mu,e_3}_\mu
\longrightarrow\Dorfman{e_1,\Dorfman{e_2,e_3}_\mu}_\mu-\Dorfman{e_2,\Dorfman{e_1,e_3}_\mu}_\mu,$$
is given by
\begin{equation}\label{defi:jacobiator}
J_{e_1,e_2,e_3} \triangleq [[A,B],C]+\mu_2(A,B)(w). \end{equation}
\end{pro}
\pf By straightforward computations, we have
\begin{eqnarray*}
  \Dorfman{\Dorfman{e_1,e_2}_\mu,e_3}_\mu&=&[[A,B],C]+\mu_0[A,B](w),
  \end{eqnarray*}
  and
  \begin{eqnarray*}
 && \Dorfman{e_1,\Dorfman{e_2,e_3}_\mu}_\mu-\Dorfman{e_2,\Dorfman{e_1,e_3}_\mu}_\mu\\&=&[A,[B,C]]+\mu_0(A)\mu_0(B)(w)-[B,[A,C]]-\mu_0(B)\mu_0(A)(w)\\
  &=&[A,[B,C]]-[B,[A,C]]+[\mu_0(A),\mu_0(B)](w).
\end{eqnarray*}
Since $[[A,B],C]=[A,[B,C]]-[B,[A,C]]$ and
$$\mu_2(A,B):\mu_0[A,B]\longrightarrow[\mu_0(A),\mu_0(B)]$$
is a natural transformation, we conclude that $J$ is also a
natural transformation. \qed\vspace{3mm}

\begin{defi}
A Dirac structure of the  omni-Lie $2$-algebra $\gl(\V)\oplus_\mu\V$
is a maximal isotropic $2$-sub-vector space (w.r.t. $\langle \cdot,
\cdot \rangle_\mu$) closed under the bracket operation
$\Courant{\cdot,\cdot}_\mu$.
\end{defi}
Similar to the proof of Proposition \ref{pro:Dirac2algebra}, the
following conclusion follows directly from Proposition
\ref{pro:jacobiator}.

\begin{pro}
Let $\LL$ be a Dirac structure of the omni-Lie $2$-algebra
$\gl(\V)\oplus_\mu\V$. Then with $J$ given in Proposition
\ref{pro:jacobiator}, $(\LL, \Courant{\cdot, \cdot}_\mu|_{\LL}, J) $
is a  Lie $2$-algebra.
\end{pro}

Given a linear functor $\huaF$ from $\V$ to $\gl(\V)$, we define a
bilinear functor,
$[\cdot,\cdot]_{\mu,\huaF}:\V\times\V\longrightarrow \V$ by
\begin{equation}\label{defi:bracketmu}
~[\xi,\eta]_{\mu,\huaF}\triangleq\mu_1(\huaF(\xi))(\eta),
\end{equation}
and a multilinear function $J:V_0\times V_0\times
V_0\longrightarrow V_1$ by
\begin{equation}\label{Jacobiator}
J_{u,v,w}\triangleq\mu_2(\huaF(u),\huaF(v))(w),\quad\forall~u,v,w\in
V_0.
\end{equation}

\begin{thm}\label{thm:twist}
With the notation given above, the graph of a linear functor $\huaF:
\V \to \gl(\V)$ is a Dirac structure of the twisted omni-Lie
$2$-algebra $\gl(\V)\oplus_\mu\V$ if and only if
$(\V,[\cdot,\cdot]_{\mu,\huaF},J)$ is a  Lie $2$-algebra.
\end{thm}

\pf For any $\xi,\eta\in\V$, we have
\begin{eqnarray*}
\pair{\huaF(\xi)+\xi,\huaF(\eta)+\eta}_\mu&=&\half\big(\mu_1(\huaF(\xi))(\eta)+\mu_1(\huaF(\eta))(\xi)\big),
\end{eqnarray*}
and
\begin{eqnarray*}
\Courant{\huaF(\xi)+\xi,\huaF(\eta)+\eta}_\mu&=&[\huaF(\xi),\huaF(\eta)]+\half\big(\mu_1(\huaF(\xi))(\eta)-\mu_1(\huaF(\eta))(\xi)\big).
\end{eqnarray*}
Therefore, similar to the proof of Theorem \ref{Thm:graphdirac},  the graph of $\huaF$  is a Dirac structure iff
\begin{eqnarray*}
\mu_1(\huaF(\xi))(\eta)&=&-\mu_1(\huaF(\eta))(\xi),\end{eqnarray*}
and
\begin{eqnarray*}
~[\huaF(\xi),\huaF(\eta)]&=&\huaF(\mu_1(\huaF(\xi))(\eta)) \equiv \huaF([\xi, \eta]_{\mu, \huaF}) .
\end{eqnarray*}
Now suppose that  the graph of $\huaF$ is a Dirac structure.
For any $u,v,w\in V_0$, we have
$$
[[u,v]_{\mu,\huaF},w]_{\mu,\huaF}=\mu_1(\huaF([u,v]_{\mu,\huaF}))(w)=\mu_1([\huaF(u),\huaF(v)])(w).
$$
On the other hand, we have
\begin{eqnarray*}
[u,[v,w]_{\mu,\huaF}]_{\mu,\huaF}-[v,[u,w]_{\mu,\huaF}]_{\mu,\huaF}
&=&\mu_1(\huaF(u))\mu_1(\huaF(v))(w)-\mu_1(\huaF(v))\mu_1(\huaF(u))(w)\\
&=&[\mu_1(\huaF(u)),\mu_1(\huaF(v))](w).
\end{eqnarray*}
Since
$$\mu_2(\huaF(u),\huaF(v)):\mu_1([\huaF(u),\huaF(v)])\longrightarrow[\mu_1(\huaF(u)),\mu_1(\huaF(v))]$$
is a natural transformation, we obtain the skew-symmetric trilinear
isomorphism
$$J_{u,v,w}:[[u,v]_{\mu,\huaF},w]_{\mu,\huaF}\longrightarrow
[u,[v,w]_{\mu,\huaF}]_{\mu,\huaF}-[v,[u,w]_{\mu,\huaF}]_{\mu,\huaF},$$
which is given by
\begin{equation}
J_{u,v,w}=\mu_2(\huaF(u),\huaF(v))(w).
\end{equation} Then the  Jacobiator identity in Definition \ref{defi:Lie2}
holds because $(\mu_0,\mu_1,\mu_2)$ is a morphism from $\gl(\V)$ to
itself. Thus $(\V, \Courant{\cdot, \cdot}_\mu, \huaF, J)$ is a Lie
$2$-algebra.

Conversely, if $(\V, \Courant{\cdot, \cdot}_\mu, \huaF, J)$ is a Lie
$2$-algebra, then
$$
\mu_1(\huaF(\xi))(\eta)+\mu_1(\huaF(\eta))(\xi)=[\xi,\eta]_{\mu,\huaF}+[\eta,\xi]_{\mu,\huaF}=0.
$$
Thus we only need to show that
$
[\huaF(\xi),\huaF(\eta)]=\huaF([\xi,\eta]_{\mu,\huaF})
$.
On one hand, we have
$$
[[\xi,\eta]_{\mu,\huaF},\gamma]_{\mu,\huaF}=\mu_1(\huaF([\xi,\eta]_{\mu,\huaF}))(\gamma).
$$
On the other hand,
$$
[\xi,[\eta,\gamma]_{\mu,\huaF}]_{\mu,\huaF}-[\eta,[\xi,\gamma]_{\mu,\huaF}]_{\mu,\huaF}=[\mu_1(\huaF(\xi)),\mu_1(\huaF(\eta))](\gamma).
$$
Since the Jacobiator is given by (\ref{Jacobiator}), we have
$$
\mu_1(\huaF([\xi,\eta]_{\mu, \huaF}))=\mu_1([\huaF(\xi),\huaF(\eta)]).
$$
Since $(\mu_0,\mu_1)$ induces an isomorphism of the underlying
$2$-vector spaces, we have
$$
\huaF([\xi,\eta]_{\mu,\huaF})=[\huaF(\xi),\huaF(\eta)],
$$
which completes the proof. \qed\vspace{3mm}

Finally, we find the corresponding Dirac structures for string type
Lie $2$-algebras.  We first construct a suitable  isomorphism from
$\gl(\V)$ to itself, where $\V$ is a $2$-vector space given by
\eqref{v} and $\gl(\V)$ is the strict Lie $2$-algebra given by
\eqref{glv}.

Since $\End(\V)$ is a 2-term DGLA,  the Lie algebra $\End^0_\dM(\V)$
represents on $\End^1(\V)$ via \eqref{representation}. Consider the
complex $(\Hom(\wedge^\bullet \End_{\dM}^0(\V), \End^1(\V)), d)$ for
the Lie algebra cohomology $H^\bullet(\End_{\dM}^0(\V),
\End^1(\V))$.  For any linear map
$\alpha:\End^0_\dM(\V)\longrightarrow \End^1(\V)$, we have the
2-cocycle
$$
d\alpha:\wedge^2\End^0_\dM(\V)\longrightarrow \End^1(\V).
$$
Define $\mu_2:\wedge^2\End^0_\dM(\V)\longrightarrow
\End^0_\dM(\V)\oplus\End^1(\V)$ by
\begin{equation} \label{eq:mu2}
\mu_2(A,B)=([A,B],d\alpha(A,B)).
\end{equation}
Then we have

\begin{lem}\label{lem:dalpha}
If $\dM:V_1\longrightarrow V_0$ is zero, then with $\mu_2$ given in
\eqref{eq:mu2},   $\mu=(\mu_0=\Id,\mu_1=\Id,\mu_2)$ is an
isomorphism from $\gl(\V)$ to itself.
\end{lem}
\pf Since $\dM=0$ the differential $\delta:\End^1(\V)\longrightarrow\End^0_\dM(\V)$ is
also zero. Thus $\mu_2$ is a bilinear natural transformation  $[\cdot, \cdot] \to [\cdot, \cdot]$.
The coherence condition \eqref{eq:higher-constraints-mu2} is equivalent to the fact that the second component of $\mu_2$ is a 2-cocycle
in $(\Hom(\wedge^\bullet \End_{\dM}^0(\V), \End^1(\V)), d)$. Since $d\alpha$ is (even exact) closed,
\eqref{eq:higher-constraints-mu2} is automatically satisfied. \qed \vspace{2mm}

 A quadratic Lie algebra is a Lie algebra $(V, [\cdot,\cdot]_V)$
together with a nondegenerate inner product $\pair{\cdot,\cdot}$ which
is invariant under the adjoint action $\ad$.
A Courant algebroid over a point is exactly a quadratic Lie algebra.
Recall from \cite{shengzhu1}, given a quadratic Lie algebra $(V,
[\cdot,\cdot]_V,\pair{\cdot,\cdot})$, there
  is an associated $2$-term $L_\infty$-algebra whose degree-$1$ part is
  $\mathbb R$, degree-$0$ part is $V$, differential $\dM$ is zero and $l_2$, $l_3$ are given
  by
  \begin{eqnarray*}
    l_2(u,v)&=&[u,v]_V,\quad\forall~ u,v\in V,\\
    l_2(u,r)&=&0,\quad \forall~ r\in\mathbb R,\\
    l_2(r,r^\prime)&=&0, \quad \forall~ r,r^\prime\in\mathbb R,\\
    l_3(u,v,w)&=&\langle[u,v]_V,w\rangle,\quad\forall~u,v,w\in V.
\end{eqnarray*}

We denote the corresponding Lie $2$-algebra by $\huaV$:
\begin{equation}\label{exLie2algebra}
\huaV=\begin{array}{c}
 \huaV_1:=V\oplus \mathbb R\\
\vcenter{\rlap{s }}~\Big\downarrow\Big\downarrow\vcenter{\rlap{t }}\\
\huaV_0:=V.
 \end{array}
 \end{equation}
The bracket functor $[\cdot,\cdot]_{\huaV}$ is given by
$$
[(u,r),(v,r^\prime)]_{\huaV}=([u,v]_V,0),
$$
and the Jacobiator $J$ is given by
$$
J_{u,v,w}=([[u,v]_V,w]_V,l_3(u,v,w)).
$$
This sort of Lie $2$-algebra is called a {\em string-type Lie
$2$-algebra} in \cite{shengzhu2} for the reason that when $V$ is a
semisimple Lie algebra equipped with its Killing form (which is
adjoint invariant), for example $\so(n)$, we arrive at the concept
of a string Lie $2$-algebra (see also Example \ref{ep:string}).

Denote by $\ad_{\huaV}:\huaV\longrightarrow\gl(\huaV)$ the induced
linear functor by the bracket functor $[\cdot,\cdot]_{\huaV}$.
Denote by $\frkG_{\ad_\huaV}$ the graph of the linear functor
$\ad_{\huaV}$.

 Evidently, $\gl(\huaV)$ is given by
 \begin{equation}\gl(\huaV)=\begin{array}{c}
 \gl(V)\oplus\mathbb R\oplus\End(V,\mathbb R)\\
\vcenter{\rlap{s }}~\Big\downarrow\Big\downarrow\vcenter{\rlap{t }}\\
 \gl(V)\oplus\mathbb R.
 \end{array}\end{equation}

Take any complement vector space $\Img(\ad)^\perp$ of the subvector space of the image of $\ad$ in $\gl(V)$.
Let $\alpha: \End_\dM^0(\huaV) =\gl(V)\oplus\mathbb R\longrightarrow \End^1(\huaV)=\End(V,\mathbb R)$
be given by
\begin{equation}\label{alpha}
  \alpha(\ad_u + X +r)(v)=\langle u,v\rangle, \quad\forall~ u \in V, X \in \Img(\ad)^\perp, r \in \mathbb{R}.
\end{equation}


By Lemma \ref{lem:dalpha},
  $\mu=(\mu_0=\Id,\mu_1=\Id,\mu_2=[\cdot,\cdot]+d \alpha)$ is an isomorphism from $\gl(\huaV)$ to
  itself. Since $\langle\cdot,\cdot\rangle$ is an invariant inner
  product on $V$, we have
  \begin{eqnarray*}
   d{\alpha}(\ad_u,\ad_v)(w)&=&[\ad_u,\alpha(\ad_v)](w)-[\ad_v,\alpha(\ad_u)](w)-\alpha([\ad_u,\ad_v])(w)\\
   &=&-\langle v,[u,w]_V\rangle+\langle
   u,[v,w]_V\rangle-\langle[u,v]_V,w\rangle\\
   &=&\langle[u,v]_V,w\rangle.
  \end{eqnarray*}
Therefore, comparing to the Jacobiator $J$ of $\huaV$, we have
  \begin{eqnarray*}
\mu_2(\ad_u,\ad_v)(w)&=&([\ad_u,\ad_v],d\alpha(\ad_u,\ad_v))(w)\\
&=&(\ad_{[u,v]_V}(w),d\alpha(\ad_u,\ad_v)(w))\\
&=&([[u,v]_V,w]_V,\langle[u,v]_V,w\rangle)\\
&=&J_{u,v,w}.
  \end{eqnarray*}
Thus by Theorem \ref{thm:twist}, we have
\begin{pro}
Let $\huaV$ be a string-type Lie $2$-algebra as in
\eqref{exLie2algebra} and
$\mu=(\mu_0=\Id,\mu_1=\Id,\mu_2=[\cdot,\cdot]+d\alpha)$ be an
isomorphism from $\gl(\huaV)$ to
  itself with $\alpha$  given by \eqref{alpha}. Then the Dirac
  structure $\frkG_{\ad_\huaV}$ of the $\mu$-twisted omni-Lie $2$-algebra
  $\gl(\huaV) \oplus_\mu \huaV$ corresponds to the string-type Lie
  $2$-algebra structure on $\huaV$ under the correspondence of Theorem \ref{thm:twist}.
\end{pro}

In particular, we realize string Lie $2$-algebras as Dirac
structures of twisted omni-Lie $2$-algebras.

\begin{ex} \label{ep:string}
We consider the Lie algebra $\so(3)$, which is isomorphic to
$\mathbb R^3$ as vector spaces. Let $e_1,e_2,e_3$ be the basis of
$\mathbb R^3$ and $\cdot$ be the canonical inner product on $\mathbb
R^3$, then the Lie bracket is given by
$$
[e_1,e_2]=\half e_3,\quad [e_2,e_3]=\half e_1,\quad [e_3,e_1]=\half
e_2.
$$
The invariant inner product $\langle \cdot,\cdot\rangle$ which is
given by Killing form turns out to be
$$
\langle e_i,e_j\rangle=e_i\cdot
e_j=\delta^i_j=\left\{\begin{array}{cc}1,& i=j,\\0,& i\neq
j.\end{array}\right.
$$
Denote the set of $3\times 3$ symmetric matrices by $\Symm(3)$.
There is a canonical decomposition
$$
\gl(3)=\so(3)\oplus \Symm(3).
$$
Since we have $\Img(\ad)=\so(3)$, we can define
 $\alpha:\so(3)\oplus \Symm(3)\oplus \mathbb R\longrightarrow \End(\mathbb R^3,\mathbb
 R)$ by
$$
\alpha(\ad_u,S,r)(v)=u\cdot
v,\quad\forall~\ad_u\in\so(3),~S\in\Symm(3),~r\in\mathbb R,~v\in
\mathbb R^3.
$$
Let  $\huaW$ be  the $2$-vector space associated to the 2-term
complex $\mathbb R \xrightarrow{0} \mathbb R^3$, and $\mu$ be the
isomorphism $\gl(\huaW)\to \gl(\huaW)$ given by
$\mu_0=\Id,\mu_1=\Id,\mu_2=[\cdot,\cdot]+d\alpha$. Then under the
correspondence of Theorem \ref{thm:twist}, the string Lie
$2$-algebra structure on $\huaW$ corresponds to the Dirac structure
$\frkG_\huaF$ of $\gl(\huaW) \oplus_\mu \huaW$, where $\frkG_\huaF$
is the graph of the functor $\huaF: \huaW \to \gl(\huaW)$ given by
\[ \huaF(w_0+w_1) (w_0'+w_1'):=[w_0, w'_0]_{\so(3)}, \quad \forall ~w_0, w_0' \in \mathbb R^3,
w_1, w'_1 \in \mathbb R.    \]

\end{ex}

\bibliographystyle{habbrv}
\bibliography{../../bib/bibz}

\end{document}